%% file: main.tex
\documentclass[]{ceurart}

\usepackage{annotation}

\usepackage[breakable,most]{tcolorbox}
\usepackage{annotation}
\usepackage{amsmath}
\usepackage{subcaption}

\usepackage{hyperref}
\usepackage{colortbl}
\definecolor{darkblue}{rgb}{0, 0, 0.5}
\definecolor{Lightgreen}{RGB}{144,238,144}
\definecolor{Gray}{gray}{0.9}

\usepackage{styles/modernbox}

\hypersetup{colorlinks=true, citecolor=darkblue, linkcolor=darkblue, urlcolor=darkblue}

\usepackage{booktabs}
\usepackage{multirow}
\usepackage{tabularx}
\usepackage{graphicx}

\begin{document}
\copyrightyear{2026}
\copyrightclause{Copyright for this paper by its authors.
  Use permitted under Creative Commons License Attribution 4.0
  International (CC BY 4.0).}

\conference{CLEF 2026 Working Notes, 21 -- 24 September 2026, Jena, Germany}

\title{Overview of the Plagiarism Detection Task at PAN 2026}
\tnotemark[1]
\tnotetext[1]{This is a corrected version of the paper published in the CLEF 2026 Working Notes (CEUR-WS Vol-4283). After publication, an error was found in the one-shot generation dataset, which contained invalid queries. The results for the one-shot dataset have been updated; the sequential dataset is unaffected. We thank Ziqian Zhang for reporting the issue. The data is available on Zenodo: \href{https://doi.org/10.5281/zenodo.23528623}{10.5281/zenodo.23528623}.}

\author[1,2]{Andr\'{e} Greiner-Petter}[%
orcid=0000-0002-5828-5497,
email=greinerpetter@gipplab.org,
url=https://gipplab.org/team/dr-andre-greiner-petter/,
]
\cormark[1]
\address[1]{Georg-August-Universit\"{a}t, G\"{o}ttingen, Germany}
\address[2]{National Institute of Informatics (NII), Tokyo, Japan}

\author[3]{Yun-Ang Wu}[%
orcid=0009-0002-3292-2571,
email=yunangwu@nii.ac.jp
]
\address[3]{Research and Development Center for Large Language Models, NII, Tokyo, Japan}

\author[2]{Yuka Kitamura}[%
orcid=0009-0004-3587-9841,
email=ykitamura@nii.ac.jp
]

\author[2,4]{Kon Woo Kim}[%
orcid=0000-0002-6970-5370,
email=ken1204@nii.ac.jp
]
\address[4]{The Graduate University for Advanced Studies (SOKENDAI), Kanagawa, Japan}

\author[5]{Maik Fr\"{o}be}[%
orcid=0000-0002-1003-981X,
email=pan@webis.de
]
\address[5]{Friedrich-Schiller-Universit\"{a}t Jena, Jena, Germany}

\author[1]{Bela Gipp}[%
orcid=0000-0001-6522-3019,
email=gipp@gipplab.org
]

\author[2]{Akiko Aizawa}[%
orcid=0000-0001-6544-5076,
email=aizawa@nii.ac.jp
]

\author[6,7,8]{Martin Potthast}[%
orcid=0000-0003-2451-0665,
email=pan@webis.de,
url=pan.webis.de
]

\address[6]{University of Kassel, Kassel, Germany}
\address[7]{hessian.ai, Darmstadt, Germany}
\address[8]{ScaDS.AI, Leipzig, Germany}

\cortext[1]{Corresponding author.}
%\fntext[1]{These authors contributed equally.}

\begin{abstract}
The generative plagiarism detection task at PAN 2026 aims at retrieving the original scientific sources for given automatically generated textual plagiarism.
We created two novel datasets of synthetically generated plagiarism following the premise of a malicious actor trying to generate papers with minimal effort while evading detection of undisclosed re-use.
We utilized four widely available LLMs: GPT-OSS 120B, Llama 3.1 8B, Qwen 3.5 27B FP8, and DeepSeek-R1 Distill Llama 8B.
In this task overview paper, we outline the creation of this dataset and summarize and compare the results of all participants with three baselines.
For this year's task, we received 26 submissions from six teams that significantly outperformed our baselines.
\end{abstract}


\maketitle
\AddAnnotationRef{}
% TODO add back in for preprint
%\AddAnnotationRef{}

\input{clef26-pan-extended-overview}

\begin{acknowledgments}
  This work has been funded by the Deutsche Forschungsgemeinschaft (DFG, German Research Foundation) -- 554559555.
\end{acknowledgments}

%% The declaration on generative AI comes in effect
%% https://ceur-ws.org/GenAI/Policy.html
\section*{Declaration on Generative AI}
The authors utilized Gemini, Claude, and ChatGPT solely for language editing and code assistance. The authors reviewed all AI-assisted content, take ultimate responsibility for the final manuscript, and confirm that any additional uses of AI are explicitly noted in the text.

%%
%% Define the bibliography file to be used
\bibliography{clef26-pan-extended-overview-lit,clef26-pan-participant-notebooks-lit}

% TODO add back in for preprint
%\input{annotation.tex}

\appendix

\section{Sequential Pipeline Prompts}

%\noindent
\begin{minipage}[t]{0.54\linewidth}
\mbox{}\\[-\baselineskip]
\begin{llmpromptbox}{Step 1: Abstract Merging}
\scriptsize
\setlength{\parskip}{0pt}

\textbf{System:} You are an expert academic writer.\\[9pt]
\textbf{Task:}
Read Abstract A and Abstract B. Write ONE merged abstract for a new survey/merged paper that integrates both sources. Keep it faithful: do not invent claims that are not supported by either abstract. If the two sources disagree, mention the disagreement neutrally. Prefer concise, information-dense writing.\\[9pt]
\textbf{Output format:} Output ONLY the merged abstract text (no title, no bullets, no JSON).\\[9pt]
\textbf{Abstract A:} \texttt{\{\{ abstract\_a \}\}}

\textbf{Abstract B:} \texttt{\{\{ abstract\_b \}\}}
\end{llmpromptbox}
\end{minipage}
\hfill
\begin{minipage}[t]{0.44\linewidth}
\mbox{}\\[-\baselineskip]
\begin{llmpromptbox}{Step 2: Abstract Compression}
\scriptsize
\setlength{\parskip}{0pt}

Rewrite the following abstract to conform to the length and tone of a formal academic abstract. Prioritize clarity, coherence, and conciseness; nonessential details may be omitted. Reduce the length to approximately one-third of the original. Do not make it sound like a combination of several works; instead, make it a coherent piece with a logical flow. Please do not write in bullet points (or a similar style).\\[9pt]
\textbf{Abstract:} \texttt{\{\{ abstract \}\}}
\end{llmpromptbox}
\end{minipage}

\begin{llmpromptbox}{Step 3: Full Paper Generation}
\scriptsize
\setlength{\parskip}{0pt}

\textbf{System:} You are an expert academic writer.

You will be given a compressed abstract for the target merged paper, Source Materials A (may be an original paper excerpt OR an already-generated paper), and Source Materials B (same).\\[9pt]
\textbf{Task:} Write a coherent merged paper that integrates A and B. Do NOT copy large chunks verbatim; synthesize and paraphrase. Do NOT invent unsupported experiments, datasets, metrics, or numeric results. If you must be uncertain, explicitly hedge. Keep style consistent across sections.\\[9pt]
Return STRICT JSON with EXACTLY the following keys:
\texttt{\{"abstract", "introduction", "methodology", "experiment", "discussion", "conclusion"\}}\\[9pt]
Each value must be a non-empty string. Output ONLY valid JSON (no markdown fences, no commentary).\\[9pt]
\textbf{Compressed Abstract:} \texttt{\{\{ abstract \}\}}

\textbf{Source Materials A:} \texttt{\{\{ materials\_a \}\}}

\textbf{Source Materials B:} \texttt{\{\{ materials\_b \}\}}
\end{llmpromptbox}

\section{One-Shot Pipeline Prompt}
\begin{llmpromptbox}{One-Shot Generation Prompt}
\scriptsize
\setlength{\parskip}{0pt}

\textbf{System:} You are an academic paper generator. You will be given \{n\} source papers. Your task: synthesize them into a single new, original research paper that draws on ideas, methods, and findings from all sources.\\[9pt]
Each section must be SUBSTANTIAL and DETAILED:
\begin{itemize}
  \item ``abstract'': 300--400 words
  \item ``introduction'': 600--800 words
  \item ``methodology'': 800--1000 words
  \item ``experiment'': 600--800 words
  \item ``discussion'': 600--800 words
  \item ``conclusion'': 300--400 words
\end{itemize}
The total paper should be at least 3500 words. Do NOT write short, superficial sections. Expand on details, provide thorough explanations, and develop arguments fully, as a real published academic paper would.\\[9pt]
Respond with ONLY a JSON object (no markdown fences, no commentary) with exactly these keys, each a non-empty string of well-written academic prose: ``abstract'', ``introduction'', ``methodology'', ``experiment'', ``discussion'', ``conclusion''\\[9pt]
\textbf{User:} Here are the \{n\} source papers to synthesize:

\texttt{[Source Paper 1]} \texttt{[Source Paper 2]} \ldots \texttt{[Source Paper n]}\\[9pt]
Now generate a single synthesized paper as a JSON object.
\end{llmpromptbox}

\section{LLM Scoring Rubric}
\label{app:scoring-prompts}

\begin{llmpromptbox}{Scorer Prompt Used for Gemini and Claude}
\scriptsize
\setlength{\parskip}{0pt}

\textbf{User:} We are selecting 50 high-quality queries from each group for the PAN 2026 generated plagiarism detection setting.\\[3pt]
\textbf{Context:}
\begin{itemize}
\setlength{\itemsep}{0pt}
\setlength{\topsep}{1pt}
\setlength{\parsep}{0pt}
    \item Each query is a generated suspicious paper intended for source detection.
    \item A good query is not just long or fluent.
    \item It should preserve meaningful retrieval signals that help recover the original source papers.
    \item At the same time, it should not be trivially easy.
    \item We have already applied a first-pass length filter.
    \item Your job is \textbf{not} to make the final selection.
    \item Your job is \textbf{only} to score each query and provide a short justification.
    \item Final selection will be done manually by a human based on your scores.
\end{itemize}
\textbf{Your task:} For each query, assign scores for the criteria below and provide a short retrieval-oriented justification.\\[3pt]

\textbf{1. Retrieval Anchor Richness} \textit{(0--2)}
\begin{itemize}
\setlength{\itemsep}{0pt}\setlength{\topsep}{1pt}\setlength{\parsep}{0pt}
    \item 0 = very few concrete retrieval signals
    \item 1 = some useful searchable signals
    \item 2 = many clear and discriminative retrieval signals
\end{itemize}

\textbf{2. Multi-Source Visibility} \textit{(0--2)}
\begin{itemize}
\setlength{\itemsep}{0pt}\setlength{\topsep}{1pt}\setlength{\parsep}{0pt}
    \item 0 = mostly looks like one dominant source
    \item 1 = some evidence of multiple source strands
    \item 2 = clear traces of at least two distinguishable source contributions
\end{itemize}

\textbf{3. Sectional Spread of Retrieval Signals} \textit{(0--2)}
\begin{itemize}
\setlength{\itemsep}{0pt}\setlength{\topsep}{1pt}\setlength{\parsep}{0pt}
    \item 0 = clues are sparse or concentrated in one small part
    \item 1 = some spread across the document
    \item 2 = useful retrieval clues are distributed across multiple sections
\end{itemize}

\textbf{4. Coherence as a Suspicious Generated Paper} \textit{(0--2)}
\begin{itemize}
\setlength{\itemsep}{0pt}\setlength{\topsep}{1pt}\setlength{\parsep}{0pt}
    \item 0 = fragmented or poorly integrated
    \item 1 = moderately coherent
    \item 2 = coherent and plausible while still preserving source traces
\end{itemize}

\textbf{5. Difficulty Balance} \textit{(0--2)}
\begin{itemize}
\setlength{\itemsep}{0pt}\setlength{\topsep}{1pt}\setlength{\parsep}{0pt}
    \item 0 = too easy or too hard
    \item 1 = somewhat balanced but imperfect
    \item 2 = good moderate difficulty for source detection
\end{itemize}

\textbf{Difficulty balance guidance:}
\begin{itemize}
\setlength{\itemsep}{0pt}\setlength{\topsep}{1pt}\setlength{\parsep}{0pt}
    \item Down-score queries that are trivially easy because of overly explicit or overly concentrated clues.
    \item Down-score queries that are too vague, too generic, or nearly unrecoverable.
    \item Prefer queries that are recoverable but still realistically challenging.
\end{itemize}

\textbf{Output format:} Query ID; Anchor Richness: X/2; Multi-Source Visibility: X/2; Sectional Spread: X/2; Coherence: X/2; Difficulty Balance: X/2; Total: X/10; Short justification: 1--3 sentences.\\[3pt]

\textbf{Important:} Do not output final keep/drop decisions. Do not rank the final top 50. Only score and justify. Keep the reasoning retrieval-oriented. Do not overvalue writing polish alone.
\end{llmpromptbox}

\input{annotation.tex}

\end{document}

%% file: clef26-pan-extended-overview.tex
\section{Generative Plagiarism Retrieval}

With the rapid advancements of large language models (LLMs), modern plagiarism detection systems must undergo a paradigm shift~\cite{WahleRFMG22}, moving away from traditional lexical-overlap approaches toward more sophisticated semantic analysis.
Modern LLMs are already capable of generating novel, highly sophisticated scientific publications that can occasionally even pass rigorous academic peer review~\cite{SiYH25,Luetal2024,Yamadaetal2025}.
Consequently, academics increasingly and legitimately use AI to proofread, polish, or expand their scientific texts.
This practice has also gained widespread acceptance within the community over the past few years.\footnote{\url{https://2023.aclweb.org/blog/ACL-2023-policy/}}\textsuperscript{,}\footnote{\url{https://neurips.cc/Conferences/2025/LLM}}

However, alongside these legitimate uses, synthetically generated articles can conceal severe cases of plagiarism and hallucinations~\cite{GuptaP25}.
Whether intentional or unintentional, these texts pose a significant risk to the integrity of the academic community.
In this landscape, the mere presence of AI-generated text does not necessarily imply academic misconduct. Hence, modern plagiarism detection systems cannot simply rely on AI classifiers.
Plagiarism detection must remain focused on identifying the uncredited reuse of ideas and text, necessitating new datasets to train and evaluate state-of-the-art approaches in a post-LLM academic landscape.

To foster research in this direction, we successfully revived the plagiarism detection task at PAN last year~\cite{GreinerPetterFWRGAP25}, providing a novel dataset featuring realistic, future-proof examples of sophisticated, generative plagiarism.
While last year's edition focused on the alignment task, i.e., identifying the exact source paragraphs for a given pair of source and plagiarized documents, this year, we shift our focus to the source retrieval subtask.

For this year's challenge, we developed two novel automated pipelines that mimic the behavior of a malicious actor.
This hypothetical actor's motivation is to generate realistic scientific publications with minimal effort while evading detection.
Our pipeline feeds well-known, widely available LLMs with multiple genuine scientific articles, instructing the models to synthesize a new, combined, plagiarized article based on those provided sources.
Participants are tasked with retrieving all sources for a given disingenuous, auto-generated article from a corpus of 100,000 scientific articles.

For this year's edition, we received a total of 26 submissions from six teams.
Four of these teams provided notebook submissions explaining their approaches~\cite{cody:2026,alkayyem:2026,wang:2026,bedratyuk:2026}, all from different countries.
Furthermore, four teams provided their source code, enabling us to evaluate their methods on future datasets and potentially varying task configurations.
In the following, we discuss the dataset creation in detail and perform a more comprehensive analysis of the submissions.
This paper is an extended version of the brief summary in the PAN 2026 overview publication~\cite{bevendorff:2026b}.

\section{Dataset}\label{sec:intro}

As in the previous year, we utilized ar5iv\footnote{\url{https://ar5iv.labs.arxiv.org/}} (an HTML5 mirror of arXiv) as the underlying source corpus for our dataset.
From this data, we constructed 4-tuples $S_k=(S_{k1},S_{k2},S_{k3},S_{k4})$ as the sources for each plagiarized document $P_k$.

\subsection{Source Tuples Selection}
\label{sec:seqential-gen}

To construct realistic, non-trivial source sets for generative plagiarism, we developed a two-stage, category-aware selection pipeline over a retrieval pool of 320,878 ar5iv document embeddings based on Specter~\cite{CohanFBDW20} (768 dimensions).
In the first stage, the pipeline selects 1,000 \textit{seed documents} to serve as anchors ($S_{k1}$) for each source set.
These seeds are sampled proportionally across arXiv categories with a per-category cap.
This constraint maximizes topical diversity across the final dataset and prevents large subject areas from dominating the sample.
In the second stage, the pipeline constructs the rest of the 4-tuple by selecting three additional companion documents ($S_{k2}, S_{k3}, S_{k4}$) for each seed.
To ensure these companion documents are highly related but do not degenerate into trivial near-duplicates, candidates are retrieved from the seed's semantic neighbors and filtered using the following criteria:
\begin{itemize}
    \item \textbf{Semantic Similarity Band:} Candidates are strictly restricted to the 60th to 95th percentile of per-source semantic similarity. This effectively prunes weakly related papers on the lower end and excludes near-duplicate matches on the upper end.
    \item \textbf{Category Overlap:} Candidates are evaluated based on their arXiv category overlap with the seed document to ensure contextual cohesion.
    \item \textbf{Overlap Penalties:} The pipeline applies penalties for excessive lexical overlap (measured via character 5-gram Jaccard similarity on titles and abstracts) and structural overlap (estimated from section-title signatures).
\end{itemize}

\begin{figure}[t]
  \centering
  \includegraphics[width=\linewidth]{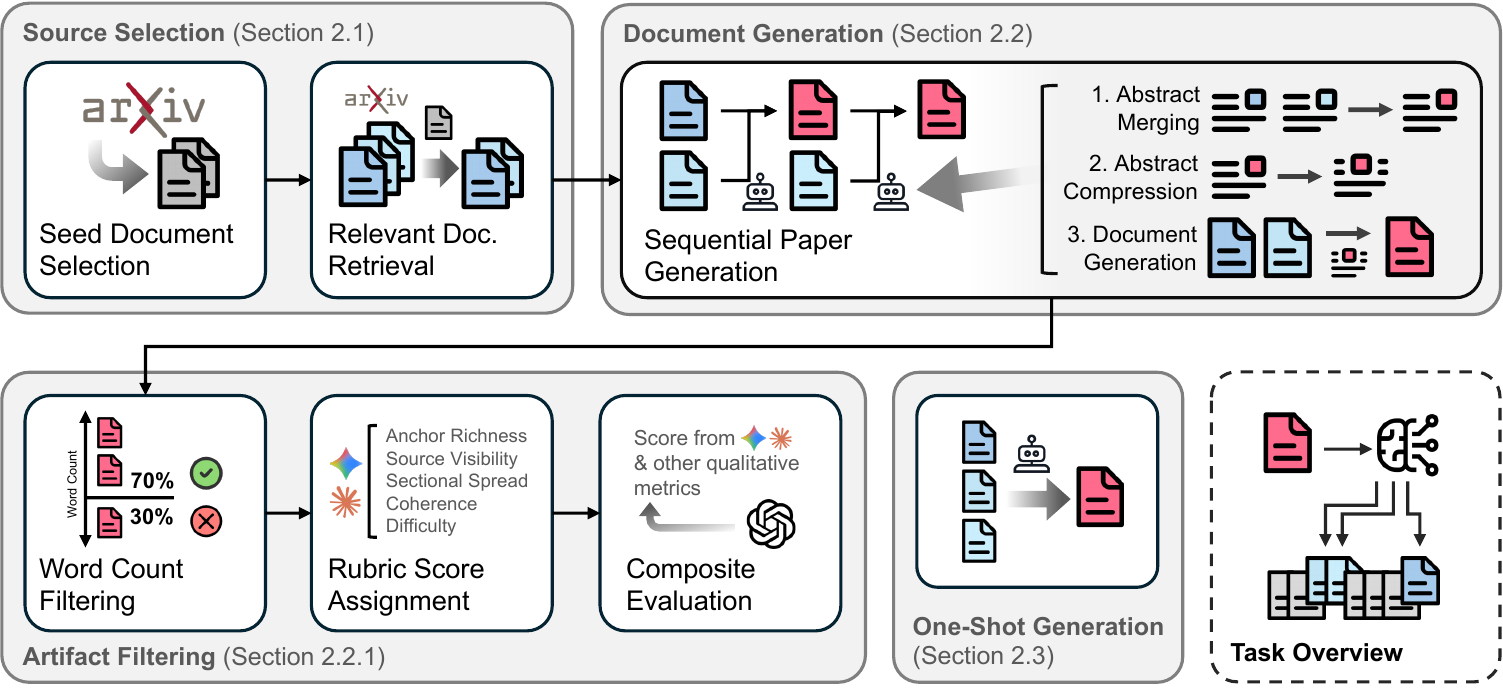}
  \caption{Overview of the dataset generation pipeline. After artifact filtering, the set of sequentially generated documents serves as the primary evaluation task data.}
\label{fig:generation_pipeline}
\end{figure}

\subsection{Sequential Plagiarism Generation}

Our primary dataset features a sequential generation pipeline.
At a high level, the sequential method merges papers via a chain: given $N$ source papers $s_1, \ldots, s_N$, it produces intermediate outputs as $(s_1, s_2) \rightarrow g_1$, $(g_1, s_3) \rightarrow g_2$, \ldots, $(g_{N-2}, s_N) \rightarrow g_{N-1}$. For this dataset, we set $N \in {2, 3, 4}$. Each merge consists of three LLM stages: abstract merging, abstract compression, and full paper generation. The final output is a JSON object with six sections: abstract, introduction, methodology, experiment, discussion, and conclusion.

The design is motivated by two technical challenges when generating a paper from multiple sources. First, models tend to merge papers in an obvious way. Since each source has a distinct narrative, the model often produces survey-like writing rather than a coherent, unified paper. To address this, we first ask the model to merge only the abstracts, as synthesizing at a higher level of abstraction makes it easier to construct a coherent argument before generating the full text. Second, we are concerned about context length: if the prompt is too long, the model may fail to produce meaningful output. By applying merges sequentially rather than all at once, we keep each prompt shorter and each individual merge higher in quality.

We split the 1{,}000 source paper sets randomly into four groups of 250, each assigned to a different model: 
%Group~1 uses 
gpt-oss-120b~\cite{openai2025gptoss120bgptoss20bmodel} (GPT-OSS), 
%Group~2 uses 
Llama-3.1-8B-Instruct~\cite{grattafiori2024llama3herdmodels} (Llama 3.1), 
%Group~3 uses 
Qwen3.5-27B-FP8~\cite{qwen3.5} (Qwen 3.5), and 
%Group~4 uses 
DeepSeek-R1-Distill-LLaMA-8B~\cite{deepseekai2025deepseekr1incentivizingreasoningcapability} (DeepSeek-R1). 
Within each group, we allocate 50, 150, and 50 instances for $N=2$, $3$, and $4$, respectively. Due to the left-fold design, earlier source papers have less direct influence on the final generated paper than later ones. We reflect this in our relevance scoring: for $N=2$ the scores are $[1, 1]$, for $N=3$ they are $[1, 1, 2]$, and for $N=4$ they are $[1, 1, 2, 3]$. Not every generation succeeds; some fail due to excessive prompt length or malformed model outputs. %Table~\ref{tab:filtering-table} reports the number of successfully generated papers per model and $N$ configuration.

This yielded 898 usable generated query artifacts across the four generator LLMs.
The remaining 102 attempts did not yield usable artifacts for subsequent filtering, most commonly because the model attempted to generate LaTeX-style text containing backslashes, which caused JSON decoding issues during the parsing stage. Specifically, 8 attempts failed for GPT-OSS, 52 for Llama 3.1, 8 for Qwen 3.5, and 34 for DeepSeek-R1.

\subsubsection{Artifact Filtering}

While the generated documents had a coherent structure, they still may contain obvious cases of copy-paste textual reuse or LLM artifacts.
Motivated by the fact that a human malicious actor would run some quick filtering in order to conceal the reuse, we also applied quality filtering.
First, we applied an LLM-specific word-count filter to the usable outputs.
Rather than imposing a single global cutoff, we selected LLM-specific thresholds to retain approximately 70\% of the usable outputs within each generator group.
The remaining queries were independently scored by Gemini 3 Flash~\cite{google2025gemini3flash} and Claude Sonnet 4.6~\cite{anthropic2026claudesonnet46} using a heuristic scoring rubric based on retrieval-anchor richness, multi-source visibility, sectional spread of retrieval clues, coherence as a suspicious generated paper, and difficulty balance. The total score was summed across all rubric categories.
GPT-5.4~\cite{openai2026gpt54} served as an automated judge for the final selection of 50 queries per LLM based on the rubric scores.
GPT-5.4 was prompted to consider the score outputs together with qualitative evidence of retrieval anchors, multi-source traces, coherence, template-like phrasing, overly explicit source cues, and obvious merge artifacts.

The final selection therefore favored queries that retained recoverable but non-trivial source traces while remaining coherent as suspicious generated papers. Candidates could be down-ranked despite favorable scorer totals when they were overly generic, too explicit, template-like, or visibly stitched together.
%The resulting benchmark contains 50 queries from each generator group, for a total of 200 queries.
The surviving queries have a mean rubric score of 8.03 and 6.05 from Gemini and Claude, respectively.
In contrast, the filtered queries have mean rubric scores of 5.89 and 4.82 from Gemini and Claude, respectively.
Table~\ref{tab:final-source-combined} provides a distribution overview of the sources across the final 200 plagiarized documents.
%
% As shown in Table~\ref{tab:scorer-summary}, the final selected queries received higher mean rubric totals from both scorer models than the non-selected candidates. However, these differences should not be interpreted as independent external validation of benchmark quality: GPT-5.4 contributed to rubric development and used the Gemini and Claude score outputs during final selection. The reported scores therefore characterize the internal LLM-assisted curation process.

\begin{table*}[t]
\centering
\caption{Source distributions for each LLM generator on the sequential dataset. The upper half shows the distribution for the number of sources for the 200 plagiarized documents. The lower half shows all (a source might have multiple categories) major arXiv categories of the sources respectively.}
\label{tab:final-source-combined}
\scriptsize
\setlength{\tabcolsep}{4pt}

\begin{tabular}{c l rrrrrr}
\toprule
& -- & GPT-OSS & Llama 3.1 & Qwen 3.5 & DeepSeek-R1 & Total & Share \\
\midrule
\multirow{3}{*}{\rotatebox[origin=c]{90}{\shortstack{Sources}}}
& 2     & 5  & 10 & 7  & 10 & 32  & 16.0\%  \\
& 3     & 30 & 27 & 35 & 30 & 122 & 61.0\%  \\
& 4     & 15 & 13 & 8  & 10 & 46  & 23.0\%  \\
\midrule
& \textit{Sources total} & 50 & 50 & 50 & 50 & 200 & 100.0\% \\
\toprule
\multirow{6}{*}{\rotatebox[origin=c]{90}{arXiv category}}
& Physics                & 33 & 30 & 32 & 37 & 132 & 66.0\% \\
& Computer Science       & 16 & 17 & 17 & 13 & 63  & 31.5\% \\
& EESS                   & 7  & 4  & 6  & 0  & 17  & 8.5\%  \\
& Quantitative Biology   & 0  & 1  & 0  & 3  & 4   & 2.0\%  \\
& Statistics             & 0  & 4  & 0  & 1  & 5   & 2.5\%  \\
& Mathematics            & 1  & 0  & 0  & 2  & 3   & 1.5\%  \\
\midrule
& \textit{Category total} & 57 & 56 & 55 & 56 & 224 & 100.0\% \\
\bottomrule
\end{tabular}
\end{table*}

\subsection{One-Shot Generation}
\label{sec:oneshot-gen}

The sequential generation pipeline, in combination with quality filtering, provides a controlled and expandable mechanism for generating plagiarism.
However, it also provides significant guardrails for the LLM to generate reused content.
Although this provides us with a relatively controlled environment to evaluate retrieval approaches, it is also significantly less realistic, as a malicious actor would not actively ensure traceable amounts of source material in the generated output.
This motivated us to create a second, considerably less restrictive dataset internally as another testing ground.

For this set, we feed $N$ source papers to the model within a single prompt, and the model is asked to generate the synthesized paper (a JSON object) in one shot.
The source papers are passed to the model directly, together with a prompt explaining the task.
The only guidelines within this prompt are aimed at length control.
Within the prompt, we include a guide specifying how many words we want for each section in order to steer the model toward generating an appropriate amount of text.

For one-shot generation, we apply the method to the same 1{,}000 source paper sets used for sequential generation.
%In contrast to the sequential method, which merges papers through a chain of intermediate steps, one-shot generation places all source papers in a single prompt and produces the synthesized paper in one LLM call, without abstract merging or compression.
Because the model must read the full text of every source and synthesize it directly, the generation step is more challenging; we therefore use only GPT-OSS, the most capable of our four selected models.
Additionally, we fix the number of input papers to $N=3$.
Since all papers contribute to the input in equal proportion, we set the relevance scores to $[1, 1, 1]$.
After removing erroneous outputs, the one-shot generation dataset contained 543 plagiarized documents.
%As with sequential generation, not every source paper set yields a successfully generated paper. 
%Table~\ref{tab:filtering-table} reports the number of successfully generated papers.

\section{Evaluation}

In the following, we evaluate all submissions and baselines using normalized discounted cumulative gain and recall, both evaluated at a specific cutoff ($\text{nDCG}@N$ and $\text{R}@N$), and mean reciprocal rank ($\text{MRR}$) without a cutoff. All metrics are averaged across the evaluation corpus.
Let $P = \{P_1, \ldots, P_{|P|}\}$ denote the complete set of plagiarized documents (queries). For a given plagiarized document $P_k$, let $D_k$ represent the ranked list of retrieved source documents returned by a system, and let $S_k$ be the ground-truth set of true source documents. The evaluation metrics are defined as follows:

\paragraph{Recall at Cutoff N (R@N)}
Recall at a specific cutoff measures the fraction of total ground-truth relevant documents successfully retrieved within the top-$N$ ranks:
$$R_k@N = \frac{|D_k[1 \dots N] \cap S_k|}{|S_k|}$$
where $D_k[1 \dots N]$ is the subset of documents found in the top-$N$ positions of the system's ranking. $\text{R}@N$ is the average of $R_k@N$ over all queries.

\paragraph{Mean Reciprocal Rank (MRR)}
Reciprocal rank targets the position of the first relevant document retrieved. If the highest-ranked true source document for a query $P_k$ occurs at rank $s_k$, its reciprocal rank is $1/s_k$ (and 0 if no relevant documents are retrieved). MRR is the average of these reciprocal ranks across all evaluation queries:
$$\text{MRR} = \frac{1}{|P|} \sum_{k = 1}^{|P|} \frac{1}{s_k}$$

\paragraph{Normalized Discounted Cumulative Gain at Cutoff N (nDCG@N)}
Discounted Cumulative Gain (DCG) accounts for the positions of all relevant documents within the top $N$ positions, penalizing relevant items if they appear lower in the ranking.
For a single query, $\text{DCG}@N$ is defined as:
$$\text{DCG}_k@N = \sum_{i=1}^{N} \frac{2^{\text{rel}_{ik}} - 1}{\log_2(i + 1)}$$
where $\text{rel}_{ik}$ is the relevance score for the retrieved document at position $i$ in $D_k$.
As the number of relevant documents varies across $P_k$ (between two and four sources), nDCG normalizes DCG by the ideal (iDCG) ordering of retrievable sources. iDCG therefore represents the maximum DCG for a given query.
$$\text{nDCG}_k@N = \frac{\text{DCG}_k@N}{\text{iDCG}_k@N}$$
As defined in Section~\ref{sec:seqential-gen}, our ideal relevance lists for the sequentially generated dataset depend on the number of sources.
For two sources, the relevance scores are $[1, 1]$; for three sources, they are $[2, 1, 1]$; and for four sources, they are $[3, 2, 1, 1]$.
For the one-shot generation dataset, the relevance scores are binary $\{0,1\}$.
$\text{nDCG}@N$ is again defined as the average over all queries.

\subsection{Baselines}

We provided three baselines: BM25, LinQ embeddings~\cite{LinqAIResearch2024}, and Numerical Matching.
The BM25 baseline ranks all documents based on their BM25 similarity to the query document.
Similarly, the LinQ baseline ranks documents based on cosine similarity, but utilizes LinQ embeddings instead of BM25 vectors.
The selection of LinQ was motivated by its status as the best-performing model in the previous year's alignment task~\cite{GreinerPetterFWRGAP25}.
Furthermore, guided by our manual analysis of several generated documents, we employed a simple numerical matching baseline.
This approach is based on the observation that generated documents likely exhibit an unusually high number of matching numerical values with their source texts, as LLMs typically treat numbers as hard facts and rarely alter them during paraphrasing.

\subsection{Team Submissions}

We received 26 submissions from six teams.
Two teams (Fsu-AIGC-Detector and NLP-DE) did not provide notebooks explaining their methodologies and are therefore listed separately in the results.
Four teams (ENGsTF, FosuAI, Bedratiuk-Lab, and Fsu-AIGC-Detector) provided code submissions, allowing us to automatically evaluate their approaches on the internal one-shot generation dataset as well.
Most teams submitted multiple runs, of which we focus only on the major increments explained by the teams.

\paragraph{Team JLP~\cite{cody:2026}}
achieved the highest scores on the sequential dataset across all metrics.
%Unfortunately, team JLP did not provide a code submission which did not allow us to evaluate their approach on the one-shot dataset.
Team JLP's system decomposes each suspicious document into sentence-level spans, using each sentence as an independent query against two complementary indices: a full-document BM25 index for lexical overlap and a dense GTE-Base v1.5 bi-encoder index over structure-aware paragraph chunks.
%Per-span BM25 and dense rankings are merged via Reciprocal Rank Fusion, then aggregated across all spans using CombSUM to produce the final source-document ranking.
A key differentiator from other teams with similar pipelines is that Team JLP constructed a 200-query synthetic development set from the source corpus, enabling systematic hyperparameter tuning via Optuna.
This allowed them to explore variables that other teams left untuned.
Most notably, span length, where single-sentence spans surprisingly outperformed all larger windows (up to 10 sentences or full-document queries).
This suggests that tighter topical focus at the sentence level is critical for isolating source-specific evidence.% in heavily paraphrased generated text.
The team submitted two runs differing only in the final aggregation step: BM25 + GTE RRF (jlp-1) sums raw RRF span scores via CombSUM, while BM25 + GTE RRF minmax (jlp-2) applies min-max normalization per span before summing.
%The normalization compresses score differences within each span, allowing weakly matched sources to contribute more evenly, which slightly improves Recall@100 but hurts precision at the top ranks, making the primary run the stronger performer on the main metrics of nDCG@10 and Recall@10.

\paragraph{Team ENGsTF~\cite{alkayyem:2026}}
employed a two-stage retrieval pipeline that is representative of several submissions.
In the first stage, source documents are split into overlapping chunks and retrieved using a hybrid dense--sparse approach.
Dense retrieval is performed either with E5 or the long-context BGE-M3 encoder, while sparse retrieval uses BM25.
Chunk-level dense scores are aggregated to the document level using max pooling and linearly combined with the BM25 scores.
In the second stage, the highest-ranked candidates are reranked with the BGE-reranker-v2-m3 cross-encoder.
Since the reranker cannot process full-length queries, the queries are chunked, and the maximum score across all query chunks is used.
The submitted runs differ primarily in the dense retriever and the use of reranking, namely E5 + FAISS, E5 + BM25, BGE-M3 + BM25, BGE-M3 + BM25 + reranker, and BGE-M3 + BM25 + reranker with score blending (bl.), where the final run linearly combines the reranker and retrieval scores instead of replacing the latter.

\paragraph{Team FosuAI~\cite{wang:2026}}
proposed a lightweight lexical retrieval approach based on Query-Chunk Provenance Modeling.
Instead of querying with the entire suspicious document, the document is split into paragraph-level chunks that independently retrieve candidate sources using BM25.
The resulting rankings are then combined through a coverage-weighted voting scheme, which favors documents that are retrieved consistently across multiple query chunks rather than through isolated lexical matches.
Unlike most other submissions, the approach relies solely on traditional lexical retrieval without dense encoders or rerankers, making it computationally lightweight while explicitly modeling multi-source provenance.
The submitted run, however, returned only the top 10 candidates, limiting the reported Recall@100 and nDCG@100 to the same values as their @10 counterparts.

\paragraph{Team Bedratiuk-Lab~\cite{bedratyuk:2026}}
proposed a two-stage retrieval pipeline combining dense retrieval with several local reranking signals.
Candidate documents are first retrieved using a MiniLM-based dense encoder.
The reranking stage combines three complementary signals.
Fragment matching aggregates sentence-level alignments with a coverage signal, while semantic attention softly aggregates similarities across multiple candidate sentences instead of relying only on the best match.
The third component compares normalized local embedding configurations to capture structural semantic correspondence beyond pairwise similarity.
The final ranking is obtained by combining the dense retrieval score with the three reranking signals, allowing the system to exploit both global document similarity and local semantic evidence.

\subsection{Discussion and Results}

\begin{table}[t]
\centering
\footnotesize
\caption{Overall results for the plagiarism detection task, grouped by team and ranked by nDCG@10 performances. 
Best results are marked in bold. FosuAI, Bedratiuk-Lab, and Fsu-AIGCC-Detector (unk-1) only retrieved 10 results and are excluded from nDCG@100 and R@100.}
\label{plagdet:table-evaluation-results}
\renewcommand{\tabcolsep}{4.5pt}
\begin{tabular}{@{}llccccc@{}}
\toprule
  \textbf{Team}     & \textbf{Approach}                     & \textbf{nDCG@10} & \textbf{nDCG@100}  & \textbf{R@10} & \textbf{R@100}     & \textbf{MRR} \\
\midrule
  JLP~\cite{cody:2026} & BM25 + GTE RRF & \textbf{0.799} & \textbf{0.833} & \textbf{0.854} & 0.963 & \textbf{0.976} \\
   & BM25 + GTE RRF minmax & 0.779 & 0.817 & 0.838 & \textbf{0.965} & 0.961 \\
  FosuAI~\cite{wang:2026}       & BM25 Segmented & 0.626 & \textcolor{Gray}{0.626} & 0.630 & \textcolor{Gray}{0.630} & 0.935 \\
  ENGsTF~\cite{alkayyem:2026}   & BGE-M3 + BM25 + RR bl. & 0.642 & 0.687 & 0.671 & 0.840 & 0.908 \\
                                & BGE-M3 + BM25 & 0.613 & 0.658 & 0.648 & 0.832 & 0.912 \\
                                & E5 + BM25 & 0.585 & 0.637 & 0.612 & 0.820 & 0.876 \\
                                & BGE-M3 + BM25 + RR & 0.491 & 0.561 & 0.565 & 0.823 & 0.695 \\
                                & E5 + FAISS & 0.385 & 0.448 & 0.433 & 0.676 & 0.647 \\
  Bedratiuk-Lab~\cite{bedratyuk:2026} & MiniLM + Reranker & 0.385 & \textcolor{Gray}{0.385} & 0.386 & \textcolor{Gray}{0.386} & 0.619 \\
\midrule
  Fsu-AIGC-Detector             & unk-1 & 0.643 & \textcolor{Gray}{0.643} & 0.657 & \textcolor{Gray}{0.657} & 0.892 \\
                                & unk-2 & 0.301 & 0.374 & 0.360 & 0.643 & 0.498 \\
  NLP-DE                        & unk-1 & 0.076 & 0.111 & 0.098 & 0.225 & 0.137 \\
                                & unk-2 & 0.040 & 0.067 & 0.057 & 0.157 & 0.076 \\
\midrule
  \rowcolor{Gray!60} Baselines  & BM25 & 0.553 & 0.607 & 0.577 & 0.782 & 0.832 \\
  \rowcolor{Gray!60}            & LinQ Embeddings & 0.391 & 0.467 & 0.445 & 0.726 & 0.699 \\
  \rowcolor{Gray!60}            & Numeral Matching & 0.184 & 0.199 & 0.170 & 0.220 & 0.321 \\
\bottomrule
\end{tabular}
\end{table}

Table~\ref{plagdet:table-evaluation-results} summarizes the results for the major approaches, ranked by the best nDCG@10 performance of each team.
Notably, BM25 proved to be a surprisingly strong baseline.
The primary reason for this robust performance likely lies in the prevalence of core scientific and mathematical disciplines within the arXiv repository.
Scientific articles in domains such as physics and computer science contain rich, highly specific terminology that often remains even after heavy paraphrasing.
These untouched, concise text fragments are ideal for TF-IDF-based retrieval methods.
Provided the texts are sufficiently long to contain multiple such concise fragments (a condition further supported by our generation and filtering pipeline), BM25 establishes a very solid foundation.
Recognizing this, several participating teams exhibited a predominant bias toward BM25 variants in their systems and often combined BM25 with a dense retrieval approach based on embeddings.
The main difference between approaches came down to decomposing queries and sources and combining the resulting scores as efficiently as possible for the deployed embedding models.

\begin{table}[ht]
\centering
\footnotesize
\caption{Overall results on the one-shot generation dataset. FosuAI, Bedratiuk-Lab, and Fsu-AIGCC-Detector (unk-1) only retrieved 10 results and are excluded from nDCG@100 and R@100. Fsu-AIGC-Detector did not submit a notebook and is listed separately.}
\label{plagdet:table-evaluation-results-oneshot}
\renewcommand{\tabcolsep}{4.5pt}
% \begin{tabular}{@{}llccccc@{}}
% \toprule
%   \textbf{Team}     & \textbf{Approach}                     & \textbf{nDCG@10} & \textbf{nDCG@100}  & \textbf{R@10} & \textbf{R@100}     & \textbf{MRR} \\
% \midrule
%   FosuAI~\cite{wang:2026} & BM25 Segmented & \textbf{0.299} & \textcolor{Gray}{0.299} & \textbf{0.286} & \textcolor{Gray}{0.286} & \textbf{0.451} \\
%   ENGsTF~\cite{alkayyem:2026} & E5 + BM25 & 0.292 & \textbf{0.311} & 0.283 & \textbf{0.344} & 0.446 \\
%          & BGE-M3 + BM25 & 0.256 & 0.277 & 0.245 & 0.320 & 0.407 \\
%          & BGE-M3 + BM25 + RR & 0.189 & 0.223 & 0.210 & 0.321 & 0.293 \\
%          & E5 + FAISS & 0.185 & 0.214 & 0.191 & 0.292 & 0.299 \\
%   Bedratiuk-Lab~\cite{bedratyuk:2026} & MiniLM + Reranker & 0.147 & \textcolor{Gray}{0.147} & 0.141 & \textcolor{Gray}{0.141} & 0.254 \\
% \midrule
%   Fsu-AIGC-Detector & unk-1 & \textbf{0.313} & \textcolor{Gray}{0.313} & \textbf{0.295} & \textcolor{Gray}{0.295} & \textbf{0.461} \\
%    & unk-2 & 0.148 & 0.180 & 0.153 & 0.263 & 0.258 \\
% \midrule
%   \rowcolor{Gray!60}Baselines & BM25 & 0.209 & 0.231 & 0.209 & 0.281 & 0.332 \\
%   \rowcolor{Gray!60} & LinQ Embeddings & 0.190 & 0.218 & 0.189 & 0.289 & 0.325 \\
%   \rowcolor{Gray!60} & Numeral Matching & 0.031 & 0.040 & 0.036 & 0.070 & 0.051 \\
% \bottomrule
% \end{tabular}
\begin{tabular}{@{}llccccc@{}}
\toprule
  \textbf{Team} & \textbf{Approach} & \textbf{nDCG@10} & \textbf{nDCG@100}  & \textbf{R@10} & \textbf{R@100}     & \textbf{MRR} \\
\midrule
  FosuAI & BM25 Segmented & \textbf{0.699} & \textcolor{Gray}{0.699} & \textbf{0.775} & \textcolor{Gray}{0.775} & \textbf{0.774} \\
  EngSTF & E5 + BM25 & 0.690 & 0.730 & 0.774 & 0.925 & 0.767 \\
   & BGE-M3 + BM25 & 0.597 & 0.645 & 0.665 & 0.859 & 0.700 \\
   & BGE-M3 + BM25 + RR & 0.447 & 0.520 & 0.585 & 0.861 & 0.503 \\
   & E5 + FAISS & 0.443 & 0.506 & 0.534 & 0.787 & 0.513 \\
  Bedratiuk-Lab & MiniLM + Reranker & 0.349 & \textcolor{Gray}{0.349} & 0.401 & \textcolor{Gray}{0.401} & 0.437 \\
\midrule
  Fsu-AIGC-Detector  & unk-1 & \textbf{0.729} & \textcolor{Gray}{0.729} & \textbf{0.797} & \textcolor{Gray}{0.797} & \textbf{0.793} \\
                     & unk-2 & 0.352 & 0.422 & 0.428 & 0.711 & 0.443 \\
\midrule
  \rowcolor{Gray!60}Baselines & BM25 & 0.477 & 0.525 & 0.550 & 0.734 & 0.570 \\
  \rowcolor{Gray!60} & LinQ Embeddings & 0.450 & 0.513 & 0.533 & 0.789 & 0.558 \\
  \rowcolor{Gray!60} & Numeral Matching & 0.065 & 0.084 & 0.087 & 0.159 & 0.087 \\
\bottomrule
\end{tabular}
\end{table}

Table~\ref{plagdet:table-evaluation-results-oneshot} provides the results of all code submissions on the one-shot generation dataset.
Team ENGsTF's best-performing approach exceeded the computing cap of 48 hours on our Tira~\cite{froebe:2023b} instance,\footnote{The instance has 5 CPU cores, 50GB RAM, and an A100 GPU with 40GB VRAM.} and is therefore not included in the results.
On this dataset, most approaches perform better except on MRR, indicating this dataset preserved more of the original documents but was more difficult to rank.
While the sequential pipeline was designed to ensure a relatively consistent level of source material within the generated document (via sequential feeding and quality filtering), the one-shot prompting approach allowed the LLM to hallucinate significantly more and choose which sources to use or not use.
As such, we would consider MRR to be the more useful metric in Table~\ref{plagdet:table-evaluation-results-oneshot}, because we cannot guarantee that all sources are still traceable within the generated document.
%On MRR, the performance drops are notably more significant. 
%The high MRR values of above 0.9, even for low nDCG or Recall scores on the sequential dataset, show that it was significantly easier to retrieve at least one source compared to the 

\begin{figure}
    \centering
    \includegraphics[width=1\linewidth]{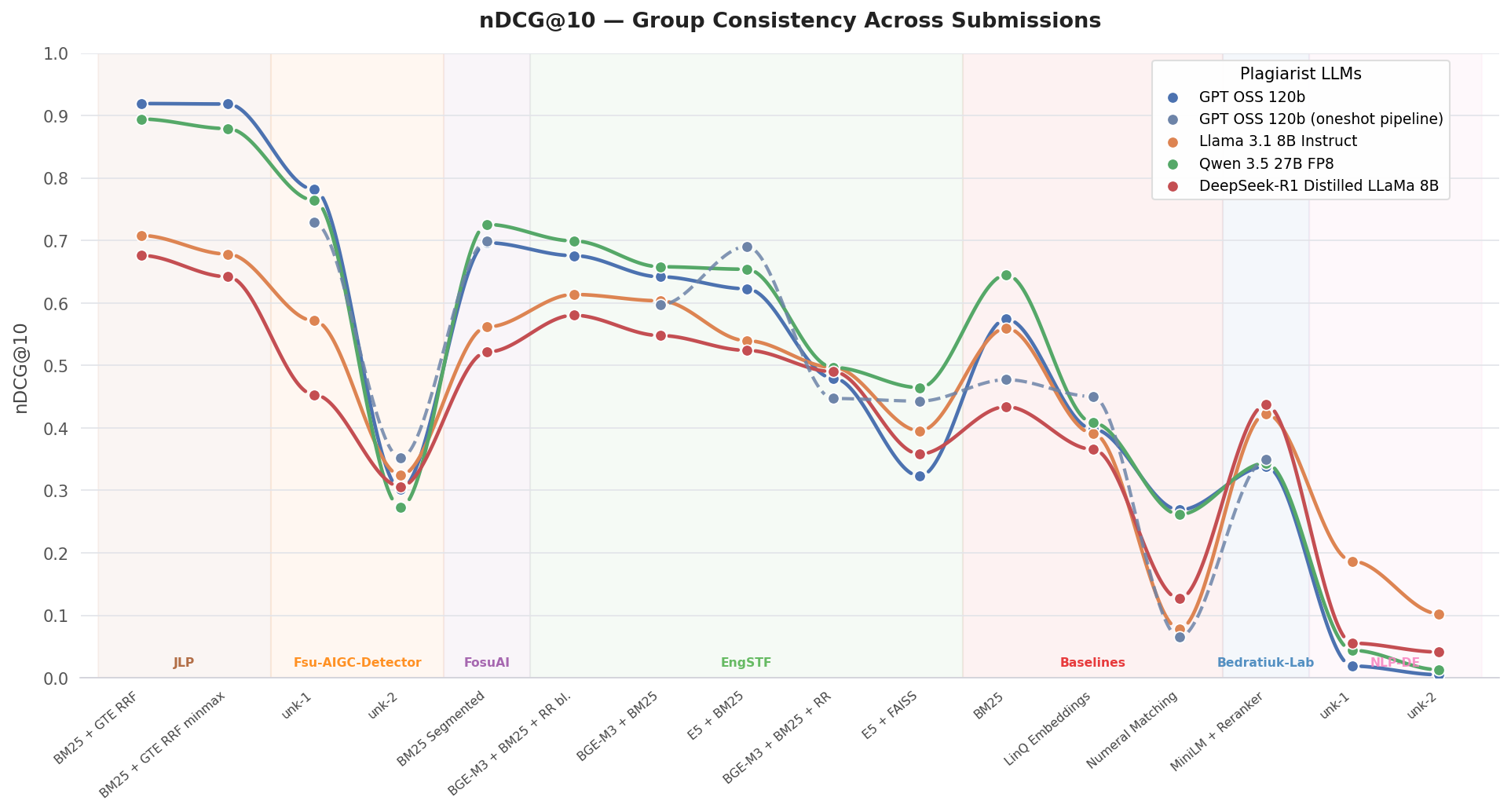}
    \caption{nDCG@10 for each generation model across all submissions and both datasets.}
    \vspace{-.2cm}
    \label{fig:ndcg-slope-llm}
    \vspace{-.5cm}
\end{figure}

\begin{figure}
    \centering
    \includegraphics[width=1\linewidth]{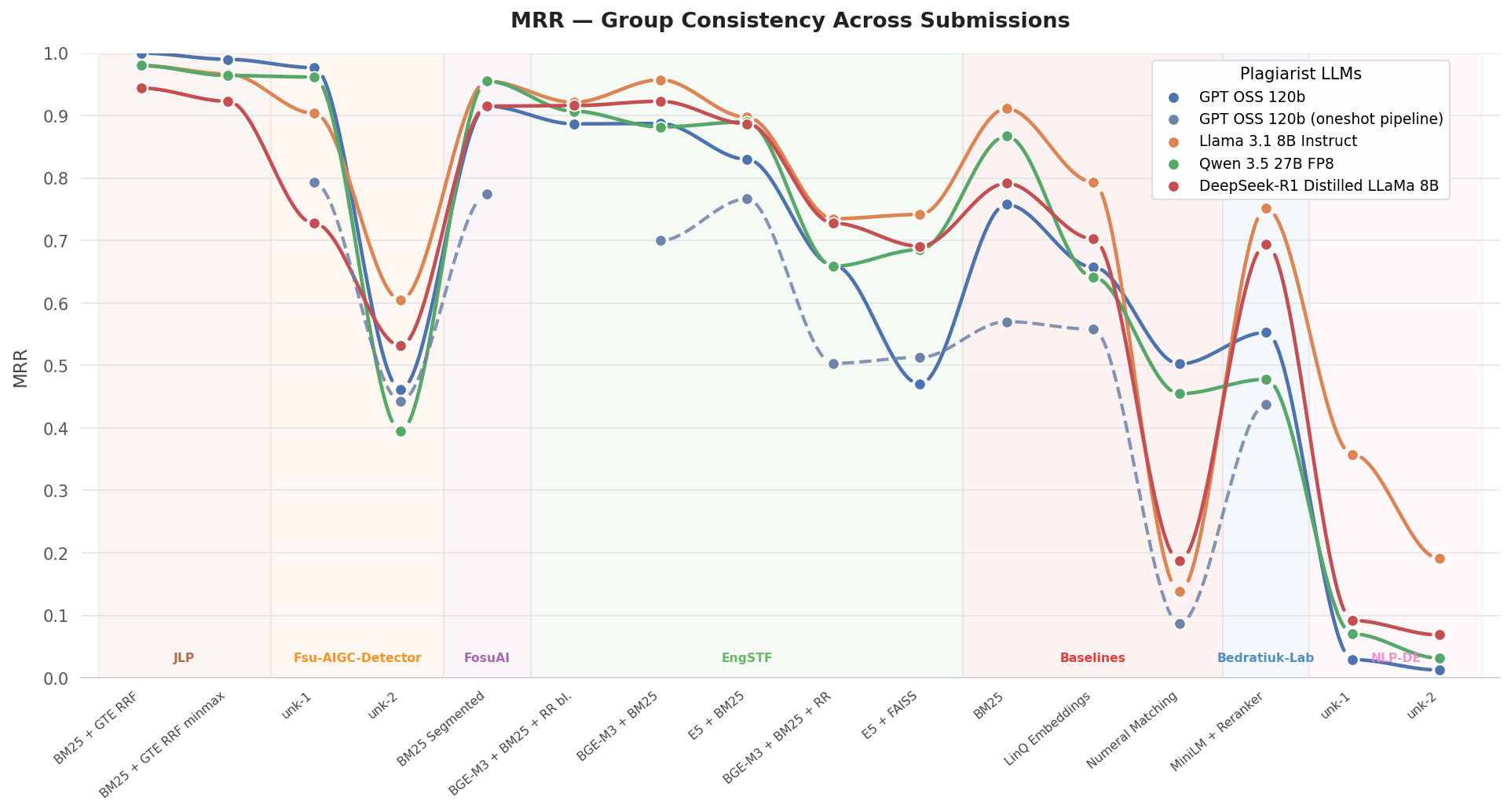}
    \caption{MRR for each generation model across all submissions and both datasets.}
    \vspace{-.2cm}
    \label{fig:mrr-slope-llm}
\end{figure}

Figure~\ref{fig:ndcg-slope-llm} shows nDCG@10 scores across all generation models for all submissions.
GPT-OSS and Qwen 3.5 sources consistently score higher in comparison to Llama 3.1 and DeepSeek-R1, with the exception of some weaker submissions.
While the proximity of Llama 3.1 and DeepSeek-R1 to each other can be explained by the selection of the distilled variant of Llama 3.1 for DeepSeek-R1, it is unclear why GPT-OSS and Qwen 3.5 exhibit similarly close performance across all submissions.
Figure~\ref{fig:mrr-slope-llm} provides the same overview but with respect to MRR scores.
Here, no clear trend across different generation models emerges.
This indicates that Llama 3.1 and DeepSeek-R1 lose more traceable source material from previous iterations during sequential feeding in comparison to GPT-OSS and Qwen 3.5.

\begin{figure}
    \centering
    \includegraphics[width=1\linewidth]{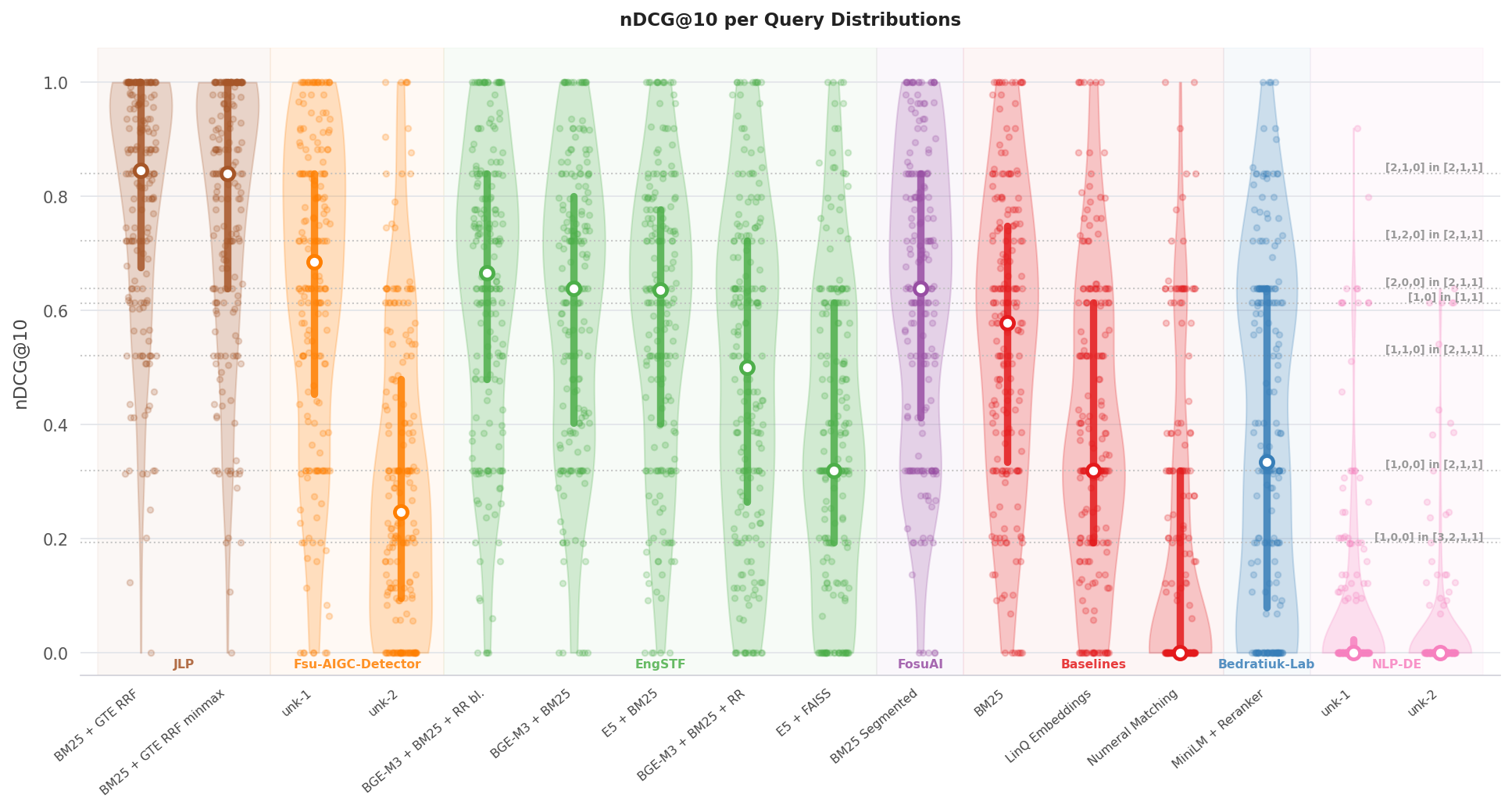}
    \caption{Distributions of all queries' nDCG@10 scores across all submissions.}
    \vspace{-.2cm}
    \label{fig:ndcg-violin}
\end{figure}

Figure~\ref{fig:ndcg-violin} further highlights the bias toward retrieving at least one source.
This figure provides the distribution of $nDCG_k@10$ scores for all queries across all submissions.
With a finite number of retrievable documents, nDCG is discrete and exhibits clear score clusters at certain retrieval combinations.
Figure~\ref{fig:ndcg-violin} highlights the most common clusters on the right.
For example, all teams except JLP and NLP-DE have a clear cluster of scores around the `$[2,0,0]$ in $[2,1,1]$' mark, i.e., they retrieved the top-relevant source (the last in the source chain) of three sources, while none of the other initial sources (the initial two documents fed to the model at the same time) were retrieved.
Team JLP's missing cluster around the same mark highlights that the initial two sources were still traceable within the plagiarized document using their approach.

\begin{figure}
    \centering
    \includegraphics[width=1\linewidth]{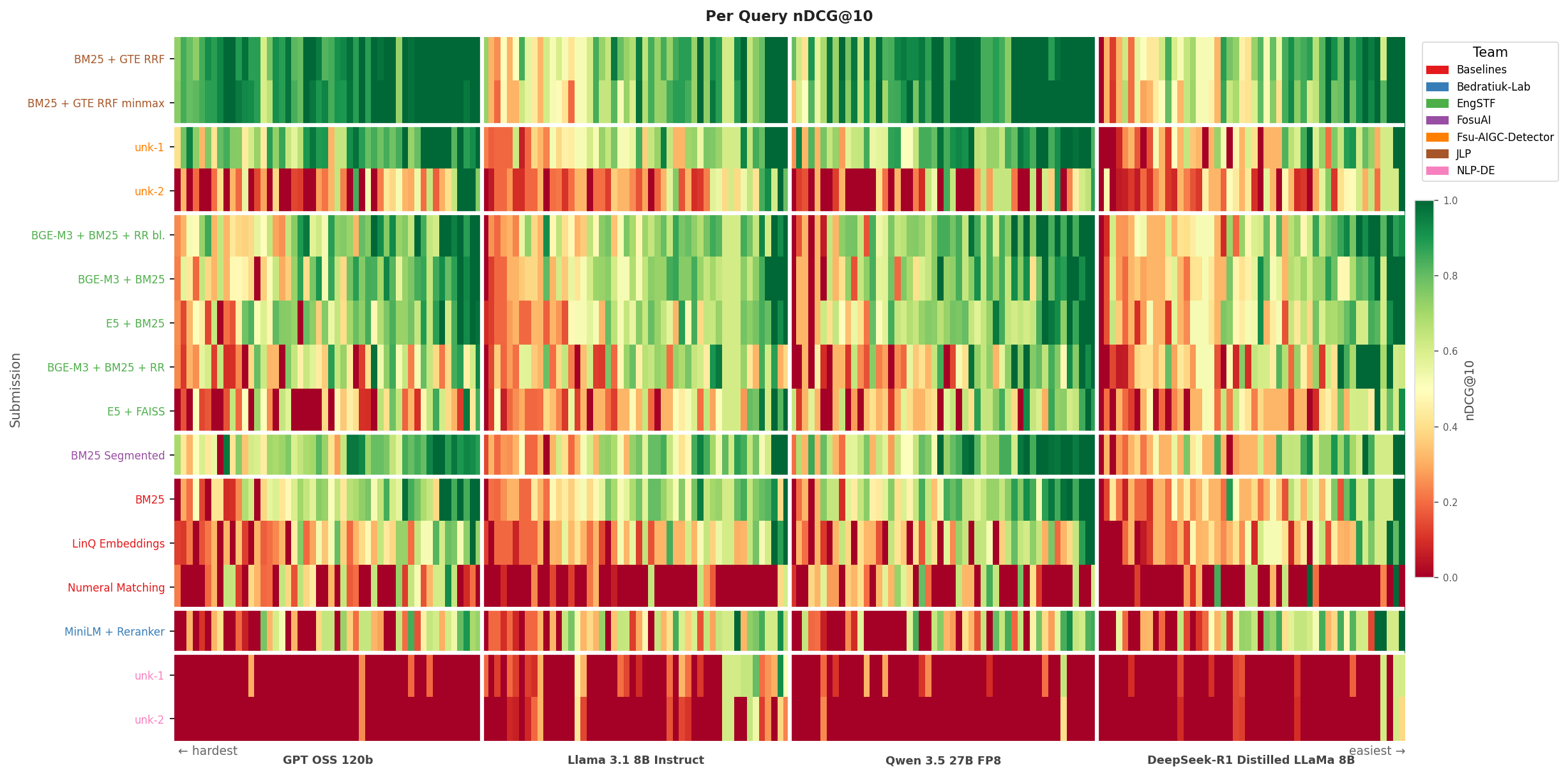}
    \vspace{-.2cm}
    \caption{Full sequential dataset nDCG@10 scores for each query and submission. The queries are grouped by the generative model and than sorted within each group by the averaged nDCG@10 over all submissions from left to right increasing. In other words, more left queries have been, the harder they were for the submitted approaches, while the more right a query appears, the more easier it has been.}
    \label{fig:heatmap}
\end{figure}

\begin{figure}
    \centering
    \includegraphics[width=1\linewidth]{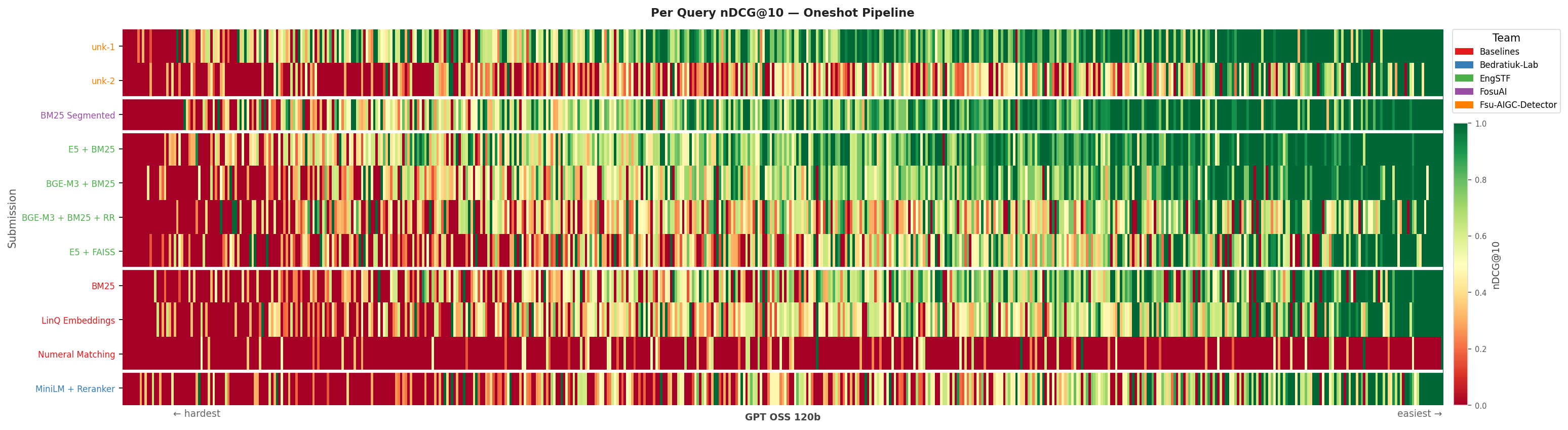}
    \vspace{-.2cm}
    \caption{Full one-shot dataset nDCG@10 scores for each query and submission. The queries are sorted by the averaged nDCG@10 over all submissions from left to right increasing. The queries achieving score 0 (left-most) are squashed.}
    \label{fig:heatmap-oneshot}
\end{figure}

Figure~\ref{fig:heatmap} provides a full overview of all 200 queries grouped by their underlying generative model and sorted from easier to harder (on average) across all submissions.
We can see that only one query, the left-most query generated by DeepSeek-R1, resulted in a score of 0 across all submissions and baselines.
This indicates that the quality measures and guardrails during generation have been largely successful in ensuring that all plagiarized documents still contain sufficient amounts of source material.
Manual investigation of this single query shows that it does indeed appear to have no obvious connections to the provided sources and instead seems to be fully hallucinated.
The query abstract focuses on digital health tools and mental well-being, while the first source discusses AI-supported story-to-video production and the second source discusses contextual vision transformers.

Figure~\ref{fig:heatmap-oneshot} provides the same overview but for the one-shot dataset.
Overall, we observe that the one-shot dataset appears to be more difficult with a larger number of queries that received low scores by all runs.
However, since the overall results improved by nDCG@10, we can assume that the spread between easy and difficult queries is wider.
This makes sense intuitively, as the missing quality controls likely generated more queries with hallucinations and no or only limited overlap with the original sources.

% Here, 396 (42.4\%) of the queries scored 0 across all submissions and baselines, i.e., no source was retrieved within the first 10 retrieved documents.
% Note that most submissions were designed to return only the top 10 results since the task's primary metrics focused mainly on the top 10 results.\footnote{For nDGC@100, 392 queries achieved a score of 0 across all submissions.}
% This further highlights the increased difficulty of the one-shot dataset but likewise also raises concerns about whether those queries could still be categorized as plagiarism.
% On the other hand, this dataset is also significantly more realistic.
% This trade-off between robust task design and realism primarily prompted us to treat the one-shot generation dataset as an internal test set for now and leave more sophisticated quality assurance for future iterations.

% TODO maybe make the heatmap for MRR (and ndcg) on oneshot?

\section{Conclusion \& Future Work}

For this year's edition, we curated two novel datasets to evaluate source retrieval methodologies for plagiarism detection systems and received 26 submissions from six teams.
Four of the six teams were able to surpass the BM25 baseline by combining BM25 with dense retrieval approaches based on model embeddings.
Team JLP submitted the winning approaches, utilizing the provided corpus to curate a synthetic development set that allowed them to fine-tune critical components of their pipeline.
% Our second dataset, based on a one-shot strategy without guardrails or quality control, was significantly more difficult for the submitted approaches.
Our second dataset, based on a one-shot strategy without guardrails or quality control, led to mixed results.
While this dataset serves as a more realistic, real-world scenario, it also raises the question of whether some generated documents still contain sufficient traceable source material to constitute plagiarism.
Without a clear answer to this question, we did not use this dataset as the primary evaluation set for this year.

%% file: annotation.tex
\clearpage
\onecolumn
\hypertarget{annotation}{}
\pagestyle{empty}
% --------- Required packages for formatting ---------
\lstset{
  basicstyle=\footnotesize\ttfamily,
  breaklines=true,
  breakatwhitespace=false,
  columns=flexible,
  numbers=none
}

% --------- Define simplified color palette ---------
% Core colors based on Tailwind blue-500
\definecolor{Primary}{RGB}{59, 130, 246}    % Main blue color (Tailwind blue-500)
\definecolor{PrimaryDark}{RGB}{30, 64, 175} % Darker blue for emphasis (Tailwind blue-800)
\definecolor{LightBg}{RGB}{239, 246, 255}   % Very light blue background (Tailwind blue-50)
\definecolor{TextDark}{RGB}{31, 41, 55}     % Dark text color (Tailwind gray-800)
\definecolor{TextMuted}{RGB}{107, 114, 128} % Secondary text color (Tailwind gray-500)

% --------- Header bar ---------
\begin{tikzpicture}[remember picture, overlay]
  \fill[Primary] ([xshift=0cm,yshift=0cm]current page.north west) rectangle ([xshift=\paperwidth,yshift=-0.4cm]current page.north west);
\end{tikzpicture}

\vspace{0.8cm}
\begin{center}
  {\fontsize{22}{26}\selectfont\sffamily\bfseries \textcolor{PrimaryDark}{CiteAssist}}\\[0.2em]
  {\Large\sffamily\scshape \textcolor{TextMuted}{Citation Sheet}}\\[0.8em]
  {\small\sffamily Generated with \href{https://citeassist.uni-goettingen.de/}{\textcolor{Primary}{\texttt{citeassist.uni-goettingen.de}}}
  % \CiteAssistCite{}
  }\end{center}

\begin{center}
\vspace{1em}
\begin{tikzpicture}
\draw[Primary, line width=0.6pt] (0,0) -- (\textwidth,0);
\end{tikzpicture}
\vspace{1.2em}
\end{center}

% --------------  BibTeX block  -----------------
\begin{tcolorbox}[enhanced,
                 frame hidden,
                 boxrule=0pt,
                 borderline west={2pt}{0pt}{Primary},
                 colback=LightBg,
                 sharp corners,
                 breakable,
                 fonttitle=\sffamily\bfseries\large,
                 coltitle=Primary,
                 title=BibTeX Entry,
                 attach title to upper={\vspace{0.2em}\par},
                 left=12pt]
\lstset{
    inputencoding = utf8,  % Input encoding
    extendedchars = true,  % Extended ASCII
    literate      =        % Support additional characters
      {á}{{\'a}}1  {é}{{\'e}}1  {í}{{\'i}}1 {ó}{{\'o}}1  {ú}{{\'u}}1
      {Á}{{\'A}}1  {É}{{\'E}}1  {Í}{{\'I}}1 {Ó}{{\'O}}1  {Ú}{{\'U}}1
      {à}{{\`a}}1  {è}{{\`e}}1  {ì}{{\`i}}1 {ò}{{\`o}}1  {ù}{{\`u}}1
      {À}{{\`A}}1  {È}{{\`E}}1  {Ì}{{\`I}}1 {Ò}{{\`O}}1  {Ù}{{\`U}}1
      {ä}{{\"a}}1  {ë}{{\"e}}1  {ï}{{\"i}}1 {ö}{{\"o}}1  {ü}{{\"u}}1
      {Ä}{{\"A}}1  {Ë}{{\"E}}1  {Ï}{{\"I}}1 {Ö}{{\"O}}1  {Ü}{{\"U}}1
      {â}{{\^a}}1  {ê}{{\^e}}1  {î}{{\^i}}1 {ô}{{\^o}}1  {û}{{\^u}}1
      {Â}{{\^A}}1  {Ê}{{\^E}}1  {Î}{{\^I}}1 {Ô}{{\^O}}1  {Û}{{\^U}}1
      {œ}{{\oe}}1  {Œ}{{\OE}}1  {æ}{{\ae}}1 {Æ}{{\AE}}1  {ß}{{\ss}}1
      {ẞ}{{\SS}}1  {ç}{{\c{c}}}1 {Ç}{{\c{C}}}1 {ø}{{\o}}1  {Ø}{{\O}}1
      {å}{{\aa}}1  {Å}{{\AA}}1  {ã}{{\~a}}1  {õ}{{\~o}}1 {Ã}{{\~A}}1
      {Õ}{{\~O}}1  {ñ}{{\~n}}1  {Ñ}{{\~N}}1  {¿}{{?\`}}1  {¡}{{!\`}}1
      {„}{\quotedblbase}1 {“}{\textquotedblleft}1 {–}{$-$}1
      {°}{{\textdegree}}1 {º}{{\textordmasculine}}1 {ª}{{\textordfeminine}}1
      {£}{{\pounds}}1  {©}{{\copyright}}1  {®}{{\textregistered}}1
      {«}{{\guillemotleft}}1  {»}{{\guillemotright}}1  {Ð}{{\DH}}1  {ð}{{\dh}}1
      {Ý}{{\'Y}}1    {ý}{{\'y}}1    {Þ}{{\TH}}1    {þ}{{\th}}1    {Ă}{{\u{A}}}1
      {ă}{{\u{a}}}1  {Ą}{{\k{A}}}1  {ą}{{\k{a}}}1  {Ć}{{\'C}}1    {ć}{{\'c}}1
      {Č}{{\v{C}}}1  {č}{{\v{c}}}1  {Ď}{{\v{D}}}1  {ď}{{\v{d}}}1  {Đ}{{\DJ}}1
      {đ}{{\dj}}1    {Ė}{{\.{E}}}1  {ė}{{\.{e}}}1  {Ę}{{\k{E}}}1  {ę}{{\k{e}}}1
      {Ě}{{\v{E}}}1  {ě}{{\v{e}}}1  {Ğ}{{\u{G}}}1  {ğ}{{\u{g}}}1  {Ĩ}{{\~I}}1
      {ĩ}{{\~\i}}1   {Į}{{\k{I}}}1  {į}{{\k{i}}}1  {İ}{{\.{I}}}1  {ı}{{\i}}1
      {Ĺ}{{\'L}}1    {ĺ}{{\'l}}1    {Ľ}{{\v{L}}}1  {ľ}{{\v{l}}}1  {Ł}{{\L{}}}1
      {ł}{{\l{}}}1   {Ń}{{\'N}}1    {ń}{{\'n}}1    {Ň}{{\v{N}}}1  {ň}{{\v{n}}}1
      {Ő}{{\H{O}}}1  {ő}{{\H{o}}}1  {Ŕ}{{\'{R}}}1  {ŕ}{{\'{r}}}1  {Ř}{{\v{R}}}1
      {ř}{{\v{r}}}1  {Ś}{{\'S}}1    {ś}{{\'s}}1    {Ş}{{\c{S}}}1  {ş}{{\c{s}}}1
      {Š}{{\v{S}}}1  {š}{{\v{s}}}1  {Ť}{{\v{T}}}1  {ť}{{\v{t}}}1  {Ũ}{{\~U}}1
      {ũ}{{\~u}}1    {Ū}{{\={U}}}1  {ū}{{\={u}}}1  {Ů}{{\r{U}}}1  {ů}{{\r{u}}}1
      {Ű}{{\H{U}}}1  {ű}{{\H{u}}}1  {Ų}{{\k{U}}}1  {ų}{{\k{u}}}1  {Ź}{{\'Z}}1
      {ź}{{\'z}}1    {Ż}{{\.Z}}1    {ż}{{\.z}}1    {Ž}{{\v{Z}}}1  {ž}{{\v{z}}}1
      % ¿ and ¡ are not correctly displayed if inconsolata font is used
      % together with the lstlisting environment. Consider typing code in
      % external files and using \lstinputlisting to display them instead.      
  }

% @inproceedings{DBLP:conf/clef/Greiner-PetterW26,
%   author={Andr{\'{e}} Greiner{-}Petter and Yun{-}Ang Wu and Yuka Kitamura and Kon Woo Kim and Maik Fr{\"{o}}be and Bela Gipp and Akiko Aizawa and Martin Potthast},
%   bibsource={dblp computer science bibliography, https://dblp.org},
%   biburl={https://dblp.org/rec/conf/clef/Greiner-PetterW26.bib},
%   booktitle={Working Notes of the Conference and Labs of the Evaluation Forum,
%                   {CLEF} 2026, Jena, Germany, 21-24 September 2026},
%   editor={Eva S{\'{a}}nchez{-}Salido and Alberto Barr{\'{o}}n{-}Cede{\~{n}}o and Alba Garc{\'{\i}}a Seco de Herrera and Sean MacAvaney and Julia Maria Stru{\ss}},
%   pages={5331--5344},
%   publisher={CEUR-WS.org},
%   series={{CEUR} Workshop Proceedings},
%   timestamp={Tue, 20 Oct 2026 17:14:40 +0200},
%   title={Overview of the Plagiarism Detection Task at {PAN} 2026},
%   volume={4283},
%   year={2026}
% }

\begin{lstlisting}
@inproceedings{GreinerPetterW26,
  author       = {Andr{\'{e}} Greiner{-}Petter and
                  Yun{-}Ang Wu and
                  Yuka Kitamura and
                  Kon Woo Kim and
                  Maik Fr{\"{o}}be and
                  Bela Gipp and
                  Akiko Aizawa and
                  Martin Potthast},
  editor       = {Eva S{\'{a}}nchez{-}Salido and
                  Alberto Barr{\'{o}}n{-}Cede{\~{n}}o and
                  Alba Garc{\'{\i}}a Seco de Herrera and
                  Sean MacAvaney and
                  Julia Maria Stru{\ss}},
  title        = {Overview of the Plagiarism Detection Task at {PAN} 2026},
  booktitle    = {Working Notes of the Conference and Labs of the Evaluation Forum,
                  {CLEF} 2026, Jena, Germany, 21-24 September 2026},
  series       = {{CEUR} Workshop Proceedings},
  volume       = {4283},
  pages        = {5331--5344},
  publisher    = {CEUR-WS.org},
  year         = {2026}
}

\end{lstlisting}
\end{tcolorbox}

% ------ Footer with subtle design element ------
\vfill
\begin{tikzpicture}
\draw[Primary!40, line width=0.4pt] (0,0) -- (\textwidth,0);
\end{tikzpicture}
\begin{center}
\small\sffamily\textcolor{TextMuted}{Generated \today}
\end{center}